\documentstyle[12pt,epsfig]{article}

\newcommand{\lsim}{\mathrel{\lower4pt\hbox{$\sim$}}
\hskip-12.5pt\raise1.6pt\hbox{$<$}\;}

\newcommand{\gsim}{\mathrel{\lower4pt\hbox{$\sim$}}
\hskip-12.5pt\raise1.6pt\hbox{$>$}\;}

\def\etal{{\it et al}.}
\def\RMP{Rev.\ Mod.\ Phys.\ }

\def\RPP{Rep.\ Prog. Phys.\ }

\def\EPC{{Eur.\ Phys.\ J.} C}

\def\NPS{Nucl.\ Phys.\ B (Proc.\ Suppl.)}
\def\NPA{Nucl.\ Phys.\ A}

\def\am{{$a_{\mu}$} }
\def\amp{{$a_{\mu}$.} }
\def\amc{{$a_{\mu}$,} }
\def\ahadc{$a_{\mu}({\rm Had;1})$, }
\def\ahadp{$a_{\mu}({\rm Had;1})$. }
\def\g2{{$(g-2)$}}

\def\noi{\noindent}
\def\bc{\begin{center} }
\def\ec{\end{center} }
\def\ahad{{$a_{\mu}({\rm Had;1})$} }
\def\ahadc{{$a_{\mu}({\rm Had;1})$,} }
\def\ahadp{{$a_{\mu}({\rm Had;1})$.} }

\def\beq{\begin{equation}}
\def\eeq#1{\label{#1}\end{equation}}
\def\eeqn{\end{equation}}

\def\beqa{\begin{eqnarray}}
\def\eeqa#1{\label{#1}\end{eqnarray}}
\def\eeqan{\end{eqnarray}}

\def\Journal#1#2#3#4{{#1} {\bf #2}, #3 (#4)}

\def\NPB{ Nucl. Phys. B}
\def\PLB{Phys. Lett.  B}
\def\PRL{Phys. Rev. Lett.}
\def\PRD{Phys. Rev. D}

\def\ZPC{Z. Phys. C}

\let\bar=\overbar

\def\etal{{\it et al.}}

\def\Dslash{\not{\hbox{\kern-4pt $D$}}}
\def\dslash{\not{\hbox{\kern-2pt $\del$}}}

\def\msb{{\bar{\ssstyle M \kern -1pt S}}}

\def\Title#1{\begin{center} {\Large {\bf #1} } \end{center}}

\begin{document}
\Title{Status of the 
Hadronic Contribution to the Muon  \g2 Value}

\begin{raggedright}  
{\it William J. Marciano \\
Department of Physics \\
Brookhaven National Laboratory \\
Upton, NY 11973 \\
Email: marciano@bnl.gov}\\
\ \\
\ \ \ {\sl and} \\
\  \\
{\it B. Lee Roberts\\
Department of Physics \\
Boston University,\\
Boston, Massachusetts 02215\\
Email: roberts@bu.edu\\ }

\bigskip\bigskip
\end{raggedright}

\bc
ABSTRACT
\ec

With the recent interest in the measured and standard model values of
the muon anomalous magnetic moment, \amc some confusion
has arisen concerning  our knowledge of the hadronic contribution 
to \amp  In the dispersion integral approach to hadronic vacuum
polarization effects, low energy contributions  must be evaluated from
data or in a model-dependent approach tested by data.  At higher 
energies perturbative QCD has been used, sometimes in conjunction with data.
The history of such evaluations is reviewed, and the prospects for 
further improvement are  discussed. We conclude that not all published
evaluations are on an equal footing or up-to-date.  
One must critically examine 
which, and how much information went into each analysis in order
to determine which are more complete, and reliable.

\vskip0.4in

\section {Introduction}

With the new result from the Muon \g2 Collaboration \cite{1999},
comparison with the standard model prediction for  \am has received
renewed scrutiny.   This result represents the third  measurement by
the E821 collaboration \cite{1998,1997}, and the first to approach the
part per million (ppm) level of precision. That level of precision
permits a stringent new test of the standard model, and a search for
physics beyond it. However, before conclusions can be drawn,  it is
necessary to have a  reliable standard model calculation with sub-ppm
accuracy for comparison with the experimental number. In this paper,
the status of  QED and weak  contributions are briefly reviewed and
then we focus on the hadronic contributions.  After an overview of all
published evaluations  of the  hadronic contribution since 1985, we
discuss several  in detail, with an eye  towards assessing both
reliability, and whether the calculation is current or has been
superseded as a point for comparison with  experiment by more
up-to-date studies. 

The standard model prediction for $a_\mu\equiv (g_\mu-2)/2$ consists of
three parts, 

\beq
a_{\mu}({\rm theory}) = a_{\mu}({\rm QED}) + a_{\mu}({\rm Hadronic}) + 
a_{\mu}({\rm Weak}).
\eeqn

Taking the value of $\alpha$ from the electron $(g-2)$ \cite{kinalpha},
yields the total QED \break contribution \cite{kinhughes,cm} 

\beq
a_{\mu}({\rm QED})=116\ 584\ 705.7(2.9)\times 10^{-11},
\eeqn

\noi which is dominated by the first-order (Schwinger) term
$\alpha/2\pi$ but is calculated (or estimated) through ${\cal
O}(\alpha/\pi)^5$. The uncertainty is very small and should not play a
role in comparisons with experiment.

The weak contribution through second order is \cite{BK1, BK2, Degr, CM2}

\beq
a_{\mu}({\rm weak }) = 152(4) \times 10^{-11} \label{eq3}
\eeqn

\noi contributing about 1.3  ppm of $a_{\mu}$ (assuming a 150 GeV Higgs
mass). The 3-loop electroweak leading logs have been estimated to
contribute $+0.5\times10^{-11}$. That small effect is safely covered by
the uncertainty in eq.~(\ref{eq3}).

Although QED and electroweak effects now appear to be well under
control,  there have been some sizeable shifts in their predicted
values over the years due to error corrections and improved higher
order calculations. In Table \ref{tab:oldvnew},  we illustrate changes
in the theoretical prediction for $a_\mu$ that have occurred since a
summary was given in 1990 \cite{kinmar}.

\begin{table}[h]
\begin{center}
\begin{tabular}{||l|c|c|c||}\hline
{\em Quantity} & {\em 1990 Value}  & {\em 2001 Value}  &
{\em Change} $(\times10^{11})$ \\
   &   $(\times10^{11})$  & $(\times10^{11})$ & \\
\hline
$a^{\rm QED}_\mu$ & 116 584 695.5(5.4) & 116 584 705.7(2.9) & $+10.2$ \\
$a^{\rm Had}_\mu$ (vac. pol 1) & 7 068(59)(164) & 6 924(62) & $-144$ \\
$a^{\rm Had}_\mu$ (vac. pol 2) & -90(5) & -100(6) & $-10$ \\
$a^{\rm Had}_\mu$ (light by light) & 49(5) & -85(25) & $-134$ \\
$a^{\rm EW}_\mu$ (1 loop) & 195(10) & 195 & 0 \\
$a^{\rm EW}_\mu$ (2 loop) & --- & $-43$(4) & $-43$ \\
\hline
 &&& \\
$a^{\rm SM}_\mu$ (total) & 116 591 918(176) & 116 591 597(67)  & $-321$ \\
\hline
\end{tabular}
\caption{Improvements in the theoretical calculation of $a_\mu$ from
1990 \cite{kinmar} to 2001. The major shifts were primarily  due to
errors in the earlier 
calculations, new calculations of higher order effects, improved
$e^+e^-\to{}$hadrons and tau data, and additional utilization of
perturbative QCD\null.}
\label{tab:oldvnew}
\end{center}
\end{table}

While the QED and weak contributions are well described perturbatively,
the hadronic contribution cannot be completely calculated from
perturbative QCD, but must instead be determined in part by using data
from the cross-section \break for \cite{B85,KNO,CLY,MD,EJ95,BW,AY}  
\beq
{e^+ e^- \rightarrow {\rm Hadrons} }
\eeqn
in conjunction with a dispersion integral. The uncertainty in those
data, particularly at low energies, largely determines the error in the
Standard Model prediction for \amp More recently, data from hadronic
$\tau$ decays have also been used along with information from perturbative
QCD at relatively low energies \cite{ADH, DH98a, DH98b}  to reduce
uncertainties.

\section{The Hadronic Contributions}

In this section we describe  the basic issues and summarize the various
evaluations of the hadronic contributions to the muon anomaly, including the 
higher-order 3-loop effects.  Then we  discuss in some detail  one of
the data-driven analyses in order to illustrate the relative importance of the
various energy regions in the evaluation of the leading hadronic
contribution,  \ahadc and the main sources
of the errors on it. Finally,  we discuss some specific details of
several of the calculations. In particular, it is hard to ignore the recent assertion
\cite{Y01} that we currently have only modest knowledge of \amc and all
calculations are of equal merit. In 
this latter section we will refute the claims made in that paper.

\subsection{The Leading Hadronic Contribution}

\begin{figure}[htb]
\begin{center}
\epsfig{file=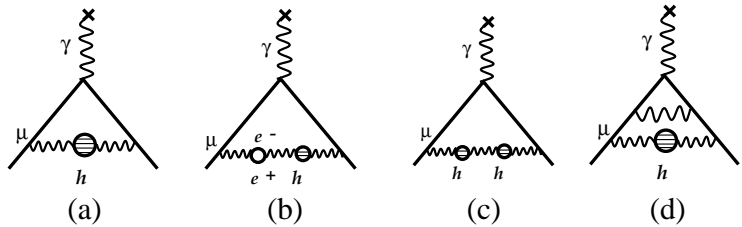 ,height=2.9cm,
}
\caption{(a) The leading  hadronic vacuum polarization contribution.
(b-d) Examples of higher 
order $(\alpha/\pi)^3$  contributions derived from the hadronic
vacuum polarization.
}
\label{fig:had}
\end{center}
\end{figure}

The leading hadronic contribution comes from the vacuum polarization diagram
shown in Fig. \ref{fig:had}(a).  Because the loop integration involves
low energy scales near the muon mass, the contributions cannot be
calculated from perturbative QCD alone.  At higher loop momenta
perturbative QCD becomes applicable, and it is common in the
evaluations of  \ahad to effectively  switch  from data to QCD at some
energy scale. 

\begin{figure}[htb]
\begin{center}
\epsfig{file=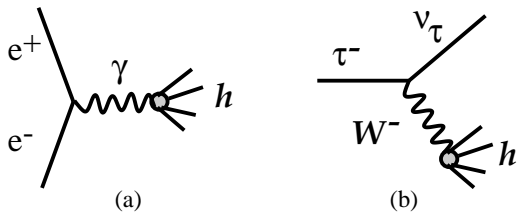,height=2.7cm}
\caption{
(a). The hadroproduction process which enters the dispersion 
relation. (b) Hadronic $\tau$ decay.
}
\label{fig:hadpro}
\end{center}
\end{figure}

Since the leading order hadronic vacuum polarization  contribution is
derived primarily from data, its value   continues to be improved by
new $e^+e^-$ cross-section measurements.  In that regard, the CMD2
collaboration at the  VEPP2M collider in Novosibirsk has collected
substantial data from the $2\pi$ threshold up  to $\sqrt{s} = 1.4$ GeV,
and is preparing a publication on these data \cite{novo}. Similarly,
the BES collaboration in Beijing has recently measured $R(s)$ from  $\sqrt{s} =
2.0 $ GeV up to 5 GeV \cite{zz}. Data from hadroproduction (see Fig.
\ref{fig:hadpro}(a)) can be related through a dispersion relation to
the first order hadronic vacuum polarization of Fig.\ref{fig:had}(a).

A second way in which the hadronic contribution can be improved is through
hadronic $\tau$ decays to vector final states. The relevant diagram is
shown in Fig. \ref{fig:hadpro}(b)  where the weak charged current can
be related to the isovector part of the electromagnetic current in
Fig. \ref{fig:hadpro}(a) through the conserved vector current (CVC)
hypothesis plus the additional requirements of isospin conservation and
the absence of second-class currents (which is the case for the
Standard Model).  While the weak $\tau$ decay proceeds through both
vector and axial vector weak-currents,  the final states with an even
number of pions (even {\sl G}-parity) are the ones relevant for \g2,
since decays to these final states go exclusively through the vector
current if there are no second class currents. Of course, isospin
violating effects of order a few percent must be properly accounted for.

The measurements of 
\beq
R\equiv{ {\sigma_{\rm tot}(e^+e^-\to{\rm hadrons})} \over
\sigma_{\rm tot}(e^+e^-\to\mu^+\mu^-)},
\eeqn
are used as input to the dispersion relation,
\beq
a_{\mu}({\rm Had;1})=({\alpha m_{\mu}\over 3\pi})^2
\int^{\infty} _{4m_{\pi}^2} {ds \over s^2}K(s)R(s)
\eeqn
 where the kernel is given by
\beq
{ K(s)\!=\! {3s \over m^2_{\mu}}\! \left\{\! x^2 (1-{x^2\over2}) 
\!+\! (1\!+\!x)^2 (1\!+\!{1\over x^2})\!
\left[ \ln (1\!+\!x)\! -\! x \!+\! {x^2\over2} \right]\! +\! 
{1+x \over 1-x} x^2 \ln x \right\} } 
\eeqn
 with
\beq
x= {1-\beta \over 1+\beta}, \qquad \beta= \sqrt{1 - {4m^2_{\mu} \over s}}.
\eeqn
 Since there is a factor of $s^{-2}$ in the dispersion relation, values
 of  $R(s)$ at low energies  dominate the calculation of 
$a_{\mu}({\rm Had;1})$. 

Here, we should point out that $R$ is often not directly measured
experimentally. In those cases where the cross-section for $e^+e^-$ is
determined using some other normalization, careful subtractions for initial
state radiation, vacuum polarization etc.\ have to be applied to the
data whereas most such effects would naturally cancel in the ratio $R$.   In
Table \ref{tab:had1} the published evaluations of \ahad  since 1985 are 
given.

\begin{table}[h!tb]
\begin{center}
\begin{tabular}{||l|l|l|l||}   \hline
{\em Ref.} &  \ahad &     \ahad  {\em in ppm}   & {\em  Comments} \\
	   & $(\times 10^{11})$ &			&   \\
\hline
B85 \cite{B85}  &  6840 (110)    & 58.6 (0.9)  &   primarily $e^+e^-$ data \\
KNO85 \cite{KNO}  &  7070 (180)    & 60.6 (1.54) & primarily $e^+e^-$ data  \\
CLY85 \cite{CLY}  &  7100 (115)    & 60.9 (0.9)  & QCD, theory and some $e^+e^-$ \\
MD90 \cite{MD}   &  7048(115)     & 60.5 (1.0)  &  $e^+e^-$ and QCD     \\
MD90 \cite{MD}   &  7052(76)      & 60.5 (0.65) &  $e^+e^-$ and QCD     \\
EJ95 \cite{EJ95} &  7024 (153)    & 60.3 (1.4)  &  primarily $e^+e^-$  data   \\
BW96 \cite{BW}   &  7026 (160)    & 60.3 (1.4)  &  primarily $e^+e^-$ data  \\
AY95 \cite{AY}   &  7113 $(103)^*$& 60.0 (0.9)  &  QCD, theory and some $e^+e^-$ \\
ADH98 \cite{ADH}  &  6950 (150)    & 59.6 (1.29) &  primarily $e^+e^-$ data  \\
\hline
ADH98 \cite{ADH}  &  7011 (94)     & 60.1 (0.8)  &     $e^+e^-$ + $\tau$ data \\
DH98a \cite{DH98a} & 6951 (75)     & 59.6 (0.6)  &$e^+e^-$, $\tau$ and
perturbative QCD  \\ 
	&		     &	           & at energies ($E>1.8$ GeV) \\
DH98b \cite{DH98b} &  6924 (62)   & 59.4 (.53)& $e^+e^-,\ \tau$ and
perturbative QCD  \\
		&&&			       and QCD sum rule constraints \\
		&&&		              at low energy \\
\hline
\end{tabular}
\caption{The first-order hadronic vacuum polarization contribution to 
\g2 obtained by a number of different authors. (Earlier evaluations
with much larger uncertainties are not exhibited.) \hfill\break
$^*$The value of \ahad given in the abstract of
\cite{AY}(hep-ph/9509378) does not agree with the value given in the
conclusions section of text.  Perhaps this confusion  was corrected in
the 1998 published version which is not available to us.  We assume 
that the value quoted in \cite{Y01} is the value to take from
\cite{AY}. Similarly, in \cite{CLY} a second method for evaluating the
$\rho$ contribution gave a somewhat smaller result 7045 which is not
illustrated in the table.
}
\label{tab:had1} 
\end{center}
\end{table}

In carrying out the dispersion integrals, there are two ways of
integrating the data.     In B85 and KNO85 the $e^+e^-$ data were fit
to models such as a Gounaris-Sakurai resonance parameterization \cite{GS}
or extensions thereof,  and these theoretical curves were then
utilized to obtain $R(s)$ for the dispersion integral. EJ95
\cite{EJ95} were the first to employ a model-independent trapezoidal
integration of data.    BW96 \cite{BW}, also used a trapezoidal integration and
took into account the correlations between the systematic errors in the
data.  Below we employ one of these two \underbar{model independent}
evaluations, which are in excellent agreement with one another,
as a benchmark to discuss other evaluations of the hadronic
contribution.  

\subsection {Higher Order Hadronic Contributions}

\begin{figure}[htb]
\begin{center}
\epsfig{file=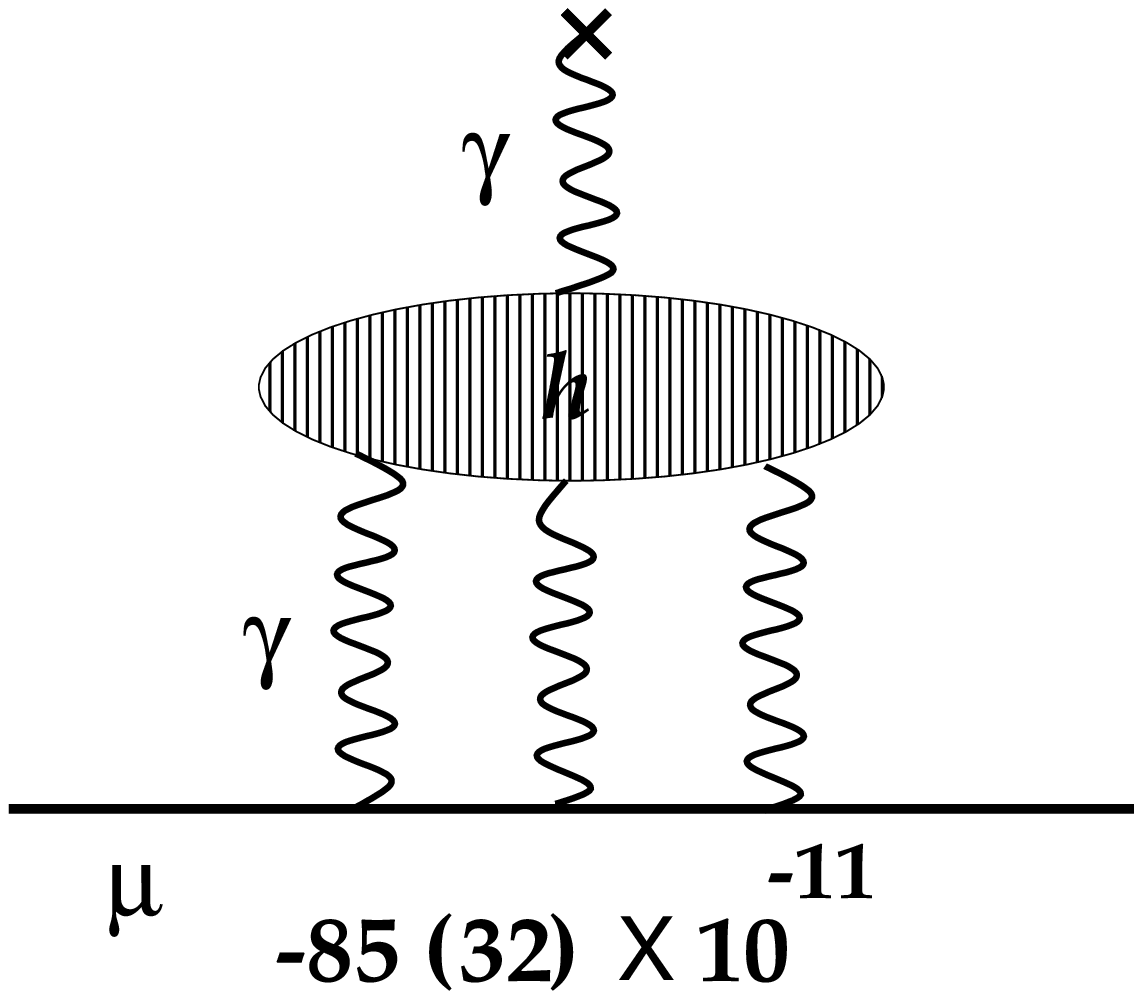,height=2.7cm}
\caption{The hadronic light-by-light scattering contribution.
}
\label{fig:hadlol}
\end{center}
\end{figure}

The higher-order (3-loop) hadronic contributions come from the hadronic vacuum
polarization  diagrams in Fig.\ref{fig:had}(b-d) and the hadronic
light-by-light scattering shown in Fig.\ref{fig:hadlol}

The higher order hadronic contributions  in Fig\ref{fig:had}(b)--(d)
were most recently evaluated by Krause \cite{krause}. The set of
diagrams represented by Fig. \ref{fig:had}(d) give a dominant negative 
contribution. Collectively, Krause found $-101(6)\times10^{-11}$. That
value was slightly updated by Alemany, Davier and H\"ocker \cite{ADH}
to 

\beq
a_{\mu}({\rm Had};2)= -100 (6) \times 10^{-11} 
\qquad (0.86 \pm 0.05) {\rm ppm}  \label{eq9}
\eeqn
which supersedes the earlier value \cite{KNO}.
The difference between  the earlier evaluation\cite{KNO} 
and the more recent calculation is attributed to
the use of the full kernel function in the new 
calculation\cite{krause}, and more up-to-date hadronic data.

\subsubsection{Hadronic Light-by-Light Contribution}

The hadronic light-by-light contribution shown in Fig.\ref{fig:hadlol}
was first calculated by KNO85 \cite{KNO}. Unlike the hadronic contributions
discussed above, this contribution cannot be determined from data, so
one is dependent on a model calculation. Einhorn \cite{ein}
pointed out that the vector meson dominance model used in KNO85 did not
satisfy the Ward-Takahashi identities which are required by
electromagnetic gauge symmetry. Furthermore, part of the
calculation was in error.

There have been three recent calculations of the
hadronic light-by-light scattering contribution, one by
Hayakawa, Kinoshita and Sanda (HKS) \cite{lol1},
one by Bijnens, Pallante and Prades (BPP) \cite{lol2},
and a follow-up improved calculation by Hayakawa and 
Kinoshita (HK) \cite{HK}.
Along the way, low energy theorems were developed by 
Hayakawa\footnote{In Ref.\ \cite{hayak} the behavior 
in the limit of low pion momentum is studied, and it
is shown that the $s$-wave component must vanish in the chiral limit under
quite general assumptions.}  for the  ($p$-wave) $V^0-\pi$ scattering
amplitude ($V^0= \rho^0,\ \omega$ or  $\phi$) \cite{hayak}. 

The situation seems to have converged as far as is possible without a
first principles or lattice calculation of the four-point  function.
Since the light-by-light contribution currently can only be obtained from
calculation, the give and take between two groups, and the mutual
checking of the other's calculation was invaluable in resolving the
magnitude and uncertainty on this correction.  In the final analysis,
it may be the uncertainty of this contribution which provides 
the ultimate limitation on the standard model prediction for
$a_{\mu}({\rm Had})$. Indeed, subtle cancellations in the calculation
deserve further study. Fortunately, the current level of uncertainty
appears suitable for the final goals of E821.

Following \cite{ADH}, we take the average of (BPP) and (HK),

\beq
a_{\mu}({\rm Had};\ {\rm lol})= -85 (25) \times 10^{-11} 
\qquad (-0.72 \pm 0.21)\ {\rm ppm} \label{eq10}
\eeqn

The total higher order hadronic correction from (\ref{eq9}) and
(\ref{eq10}) is
$$ a_{\mu}({\rm Had;\ Higher\ order})=  -185 \pm 26 \times 10^{-11}   $$

\section{The Hadronic Contribution Compared with \boldmath$a^{\rm exp}_{\mu}$}

\begin{figure}[htb]
\begin{center}
\epsfig{file=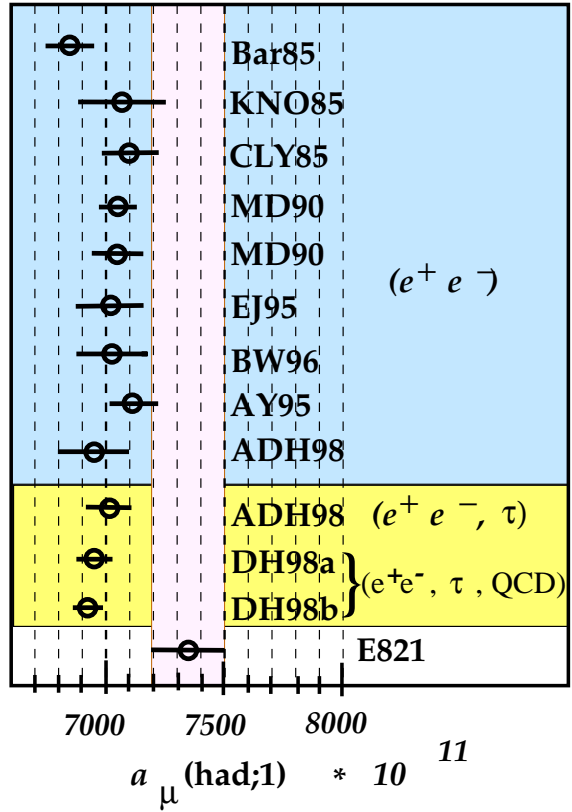,height=10.7cm}
\caption{A comparison of the calculated values of \ahad along with
the difference between the measured value of \am less the 
QED, weak and higher order hadronic corrections.
}
\label{fig:hadgraph}
\end{center}
\end{figure}

The recent result reported by E821 is
\beq
a^{\rm exp}_{\mu} = (116\ 592\ 020 \pm 160) \times 10^{-11}
\eeqn
where the systematic and statistical errors have been added in quadrature.
When combined with the previous measurements, one gets
\beq
<a_{\mu}>_{\rm exp} = (116\ 592\ 023 \pm 151 )  \times 10^{-11}
\eeqn

\noi Since the QED and electroweak contributions are not in question, we
can subtract those calculated values from the experimental number.
The resulting quantity represents the hadronic contribution plus any
contribution from new physics.  Next, we subtract off the
higher order hadronic contributions given above and obtain

\beq
 a_{\mu}({\rm Had;1}) +  a_{\mu}({\rm New?}) =  7350 (153) \times 10^{-11}.
\eeqn

\noi This number is to be compared with the values of \ahad illustrated
above in Table \ref{tab:had1}.
A graphical comparison is given in Fig. \ref{fig:hadgraph}.

This compilation represents a completely uncritical selection of published
evaluations.  Nevertheless, all evaluations agree with each other and
deviate in varying degrees with the value suggested by experiment.  We discuss
some of these in detail below.  Because of the common data sets which
went into most of  these evaluations, the errors on these 
evaluations of \ahad are highly correlated.

\section{The Different Evaluations of \boldmath\ahad}

We now give a detailed critique of the evaluations of \ahadp Rather
than starting the discussion with 
 B85, KNO85, or CLY85, all  of which used a parameterization 
for the $\rho$ resonance   we first consider the 
model-independent analyses of EJ95 and BW96.

While the analyses of $e^+e^-$ data by EJ95, BW96 and ADH98 
obtain essentially  identical final results, 
there are differences on several points between the three analyses.  
In EJ95 two results are quoted. These are given in Table 3a, and Table 3b of
ref. \cite{EJ95}.
It is the former which is directly comparable to the other analyses.
The so-called 
``renormalization group improvement'' result quoted in EJ95 (Table 3b) 
represents an effort by those authors to incorporate the higher order
contribution of Fig. \ref{fig:had}(b) into the lowest order hadronic vacuum
polarization. Unfortunately, that approach is invalid and should be
discounted. The fact that it is quoted in the abstract of EJ has caused
some recent confusion \cite{Y01}. The authors (EJ) themselves have not
continued to advocate the  Renormalization Group improved approach, but
instead have accepted the conventional method of  computing
higher order hadronic corrections separately as done by Krause.\cite{krause}

Although both EJ95 and BW96 do a model independent analysis,
there is a difference in how errors and experimental data are combined.
Up to 2 GeV, the data have traditionally been published as exclusive cross
sections.  Above 2 GeV they are published as the inclusive cross section
ratio $R$(s).  It is necessary to integrate over energy, sum over modes (below
2 GeV), and
combine results from different detectors.   EJ95
computed an error weighted average of $R$(s) over all experiments, but in so
doing, they lost the information on correlations between the errors
over energy. 

BW96 combined the data in a way which permitted the correlations
to be included, and they also added a scale
factor when combining results from different experiments, thereby handling
experiments which do not agree in a conservative manner.

While the EJ95 method gives up the information on the 
correlations over energy, and BW96
explicitly includes them, the final results are so close that 
there is no practical difference between the two methods or results.
We conclude that the correlations are not important.  Furthermore,
when the better data which are now available from BES and other
$e^+e^-$ facilities are included in
the future, the issue becomes irrelevant.

\subsection{The Contributions to \boldmath$a_{\mu}({\rm Had};1)$ from Different Energy
Regions}

It is important to understand the source of the  errors on
$a_{\mu}({\rm Had};1)$ how they have recently been improved and  will be
further improved
over time as new data become available.
In Table \ref{tab:had1bw}, the 
contributions to
$a_{\mu}({\rm Had};1)$ from different energy regions in the dispersion
integral  found in 1996 by Brown and Worstell 
are listed with their uncertainties.  

\begin{table}[htb]
\begin{center}
\begin{tabular}{||l|l|l|l|l||}   
\hline
Energy Region & $a_{\mu}({\rm Had};1)$ & \% of & $\delta_{\rm tot}$ &
 $\delta_{\rm frac}$ \\
(GeV) & ($\times 10^{11} $) & $a_{\mu}({\rm Had};1)$ & ($\times 10^{11}$) 
& (ppm) \\ 
\hline
$\sigma(e^+e^-\rightarrow {\rm Hadrons})$ $\sqrt s <$ 1.4 & 6113.32 & 87 
& 149.97 & 1.29 \\
$\sigma(e^+e^-\rightarrow {\rm Hadrons})$ $1.4 \leq \sqrt s \leq 2.0$ &
324.66 & 4.6& 24.96 & 0.21 \\ 
\hline
$R(s)^a$ $2.0 \leq \sqrt s \leq 3.1$ & 283.74 & 4.0 & 35.51 & 0.30\\ 
$R(s)^b$ $2.0 \leq \sqrt s \leq 2.6$ & 204.8 &2.9& 31.47 & 0.27 \\
$R(s)^b$ $2.6 \leq \sqrt s \leq 3.1$ & 78.93 &1.1& 13.99 & 0.12\\
\hline
$J/\psi$ (6 states) & 90.47 & 1.3 & 9.69 & 0.08\\
$\Upsilon$ (6 states) & 1.09 &-&  0.13 & -  \\
QCD $3.1 \leq \sqrt s < \infty $ & 213.01 & 3.0 & 3.71 & 0.03 \\
\hline
Subtotal \ \ $\sqrt s <$ 3.1 $+\ J/\psi\ +\ \Upsilon$
(no QCD)
& 6813.28 & 97 & 160.22 & 1.37 \\
Subtotal\ \ \  $\sqrt s <$ 1.4 \qquad \qquad \qquad & 6113.32 & 87 &
149.97 & 1.29 \\
Subtotal 1.4 $< \sqrt s <$ 3.1 \qquad \qquad & 608.40 & 8.7
& 55.51 & 0.48 \\
Subtotal\ \  $\sqrt s >$ 1.4 \ \ 
(Includes QCD)
 & 912.97 &13&
56.48 & 0.48 \\
\hline
Total {had1} & 7026.29 & 100 & 160.25 & 1.37\\
\hline
\end{tabular}
  \caption{
  The 1996 contributions to $a_{\mu}({\rm Had};1)$ from the various energy 
   regions with their total errors.  The systematic errors are twice the
statistical errors in the region from threshold to 2.0 GeV
and from 2.6 to 3.1 GeV\null.  From 2.0
to 2.6 GeV the systematic and statistical errors are about equal.
(This table is based on
   Table IX. from Ref. \cite{BW}). } 
\label{tab:had1bw} 
\end{center}
\end{table}

The largest contribution comes from threshold to 1.4 GeV.  However,
in the context of this model-independent analysis, one sees that
if the error from the $2\pi$
threshold to 1.4 GeV were eliminated completely, one would still have been
left with an error of $\delta_{\rm tot} = 56.5 \times 10^{-11}$, or
0.48 ppm.   The recent Aleph $\tau$-decay data have been used
to significantly improve the region below 1.4 
GeV\null, and inclusion of the full LEP and CLEO $\tau$ data samples
along with the new  $e^+e^-$ Novosibirsk data will improve things further.
 
The region between 2.0 and 2.6 GeV had been particularly problematic.
In 1996 there were only two experiments, BCF and
$\gamma\gamma2$, which have comparable total errors, but $\gamma\gamma2$ has a
large systematic error.   This small region alone
contributed an uncertainty of $\sim 0.27$ ppm to $a_{\mu}({\rm Had};1)$, while
the entire region from 2.0 to 3.1 GeV contributed $\sim 0.31$ ppm error. 
(Without correlations these errors add in quadrature.) 
With the new data from BES \cite{zz}, which agree  with the QCD 
evaluation by Davier and H\"ocker, it appears that the problems in this
energy region are solved and the uncertainty is significantly reduced.
Hence, more current studies of \ahad have justifiably smaller
uncertainties than earlier efforts.

\subsection{Brief Overview of the Evaluations of \boldmath$a_{\mu}({\rm Had};1)$}

Before giving a more detailed discussion of the evaluations listed in
Table \ref{tab:had1}, we provide a brief perspective.  
Our point of view is that
the knowledge of the hadronic contribution to $a_\mu$ is an evolving
topic, that 
earlier analyses which represented the state of the art at
one time, become outdated when new data and improved evaluations become available.

When B85 and KNO85 made their analyses, there were many unpublished 
data from Orsay which were not included.  By 1995 these data were 
mostly published, so we believe EJ95, BW96 and ADH98 supersede
the earlier studies. We defer the discussion of CLY85
and AY95 to the next section, but note in passing
that one of the goals of MD90 was to improve the estimate of
the theoretical errors presented in CLY85.

The values of \ahad obtained by
EJ95, BW96 and ADH98 are quite consistent.  It is interesting to note
the large improvement obtained by ADH98 by the inclusion of the 
$\tau$-decay data.  From Fig. \ref{fig:hadgraph} and Table \ref{tab:had1}
one can see that the addition of the $\tau$-decay data raised the
value of \ahad slightly, but to a value quite consistent with
the earlier analyses.  The use of QCD for 
$\sqrt{s}>1.8$ GeV by DH98a, along with
$e^+e^-$ and $\tau$-decay data lowered the value of \ahad and
significanly reduced the error,  since
the QCD prediction was systematically below the existing data
(discussed above) in the energy region 2-3 GeV.  
This QCD prediction, which was
done \underbar{in advance} 
of the recently reported $R(s)$ measurement at 
Beijing \cite{zz}, is in excellent agreement with the new data,
which do not seem to suffer from the systematic 
problems of the older BCF and
$\gamma \gamma 2$ data \cite{H01}.

This excellent agreement between the QCD calculation and the new
$R(s)$ data gives one confidence in the evaluation of \ahad
presented in DH98a.  Since the input to this evaluation 
contains much more data than the earlier evaluations, and the 
theory input seems to be justified by the new Beijing data, we
believe that one must at least take DH98a as the best evaluation of
\ahad up to that point.  In DH98b,  the same authors use QCD sum rule constraints
at lower energies to further improve the uncertainty on their
evaluation of \ahadc and it is this last value which is used in
Ref.\ \cite{1999} to 
compare the experimental and theoretical evaluations. That final
improvement is perhaps a more controversial improvement. However, we
note that DH98a and DH98b do not significantly differ.

To illustrate the degree of improvement in \ahadc we give in Table~4
contributions from different energy regions as obtained by Davier and
H\"ocker.\cite{DH98b}
\begin{table}[h]
\begin{center}
\begin{tabular}{||l|c||}\hline
$\sqrt{s}$ (GeV) & $\mbox{\ahad} \times10^{11}$ \\
\hline
$2m_\pi-1.8$ & $6343\pm60$ \\
1.8--3.7 & $338.7\pm4.6$ \\
$3.7\mbox{--}5 + \psi(1s,2s)$ & $143.1\pm5.4$ \\
5--9.3 & $68.7\pm1.1$ \\
9.3--12 & $12.1\pm0.5$ \\
12--$\infty$ & $18.0\pm0.1$ \\ \hline
Total & $6294\pm62$ \\
\hline
\end{tabular}
\caption{Contributions to \ahad from different energy regions as found
by Davier and H\"ocker.\cite{DH98b} \label{tab:DH}}
\end{center}
\end{table}
Although the energy divisions in Tables~3 and 4 do not
coincide, one can see indications of improvement throughtout Table~4.
Particularly significant is the error reductions for
$2m_\pi\le\sqrt{s}\le1.8$ GeV due to tau data as well as for $\sqrt{s}
\gsim1.8$ GeV due to the use of perturbative QCD (now confirmed by BES
data).

\subsection{Detailed Discussion of CLY85 and AY}

We now discuss some of the specific issues raised by Ref.\ \cite{Y01}.
Before undertaking this discussion, we note that the
evaluations of \ahad in CLY85 and AY95 are somewhat high, but
consistent with other 
analyses, and for some of the same reasons that B85, KNO85, EJ95, etc.\
are now considered to be obsolete, these two should also be so considered.
We also recognize that CLY85 represented the first attempt to use
perturbative QCD down to low energies in the evaluation of \ahadp
While it took some time to be accepted, 
the use of QCD, along with $\tau$-decay data has resulted in substantial
improvements \cite{DH98a,DH98b}. Nevertheless, we feel compelled by the
confusion which has been recently  
generated by Ref.\ \cite{Y01} to detail why we feel that CLY85 and AY95
should not be used in comparison with experiment, at least
not with the same conviction as later improved determinations.  In
fact, we argue that both should be updated before a serious comparison
with more up-to-date approaches can be made.

Our first criticism of CLY85 and AY95 involves their treatment of the
$\rho$ resonance. Both employ a parametrization of the pion
electromagnetic form factor based on Particle Data Group $\rho$ meson
parameters, rather than direct data, and 
they assign a very small uncertainty
to their evaluation of the dispersion integral in that important
low-energy region. Comparison of their results with more recent
data-driven studies suggest that their evaluation of the $\rho$ 
contribution
was too large. Part of the problem may be traced to their use of a
relatively low mass for the $\rho$, 768.5 MeV, and larger width
$\Gamma_\rho\simeq150$ MeV when compared with more recent detailed
studies of $e^+e^-$ and tau decay data which suggest $m_\rho\simeq776$
MeV and $\Gamma_\rho\simeq146$ MeV\null. At the very least, the
uncertainty inherent in their approach should have been considerably larger.
However,  if one merely changes the $\rho$ mass and width to the values
currently more consistent with data, the AY value of $a_\mu({\rm
Had};1)$  is (very) roughly lowered by $\sim 100$, ($\times 10^{-11}$) a 
significant reduction. To put
it on a level with more data-driven recent studies, it would need
to be fully updated. 

\begin{figure}[h!tb]
\begin{center}
\epsfig{file=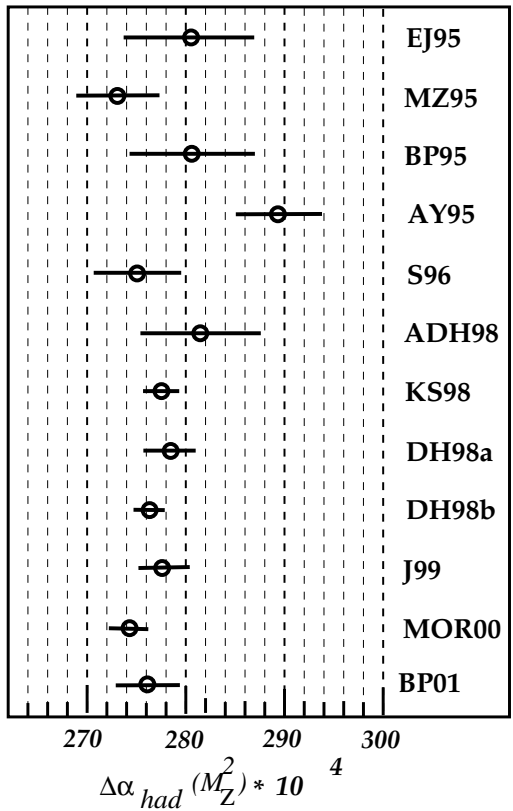,height=10.7cm}
\caption{A comparison of the calculated values of 
$\Delta \alpha_{had}(M_Z^2)$.
}
\label{fig:alphagraph}
\end{center}
\end{figure}

A second more disturbing problem with the AY95 analysis is its use of
perturbative QCD down to very low energies $\sim 1.4$ GeV and its
unconventional 
treatment of heavy quark thresholds used in conjunction with resonance
contributions evaluated by other authors. Rather than comment on the details
of their analysis, we consider their results for another dispersion
integral,
\beq
\Delta \alpha_{\rm had} =
-{ \alpha M^2_Z \over 3 \pi} {\rm Re} 
\int_{4 m^2_{\pi}}^\infty ds {R(s) \over s(s-M_z^2)-i \epsilon}\ ,
\eeqn
the hadronic loop corrections to the fine structure constant
$\Delta\alpha^{(5)}_{\rm had}(m^2_Z)$. That quantity has been evaluated
by many 
authors\cite{EJ95, ADH, DH98a, DH98b, MZ95, BP95, S96, KS98, J99, MOR00, BP01}
 because of its critical use in comparing $\alpha$,
$m_Z$ and $G_\mu$ with precision measurements of $m_W$,
$\sin^2\theta^{\rm eff}_W$, etc. In Fig.\ref{fig:alphagraph}, 
we compare various
evaluations of that quantity. Notice, its significant improvement in
the more recent evaluations primarily because of improved data. The
only calculation which deviates significantly from 
the others is the value
\beq
\Delta\alpha^{(5)}_{\rm had}(m^2_Z) = 289.4\pm4.4\times10^{-4} 
 \qquad  ({\rm AY95}) \label{eq14}
\eeqn
 given in the text of Adel and Yndurain \cite{AY}. (As in the case of
hadronic contributions to $a_\mu$, AY95 give very different results for
$\Delta \alpha^{(5)}_{\rm had}(m^2_Z)$, called $\Delta\alpha_h$, in
their text and abstract.)

The deviation of Eq.\ref{eq14} by about $3\sigma$ from
the central value of more up-to-date
evaluations, which give $\sim276$--$277\times10^{-4}$ would seem to invalidate
the perturbative analysis of heavy quark thresholds used in conjunction
with explicit resonance contributions for that quantity (suggesting
double counting)
and thereby call into question its use for $g_\mu-2$ studies. We might
note that comparison of $m_W$ and $\sin^2\theta^{\rm eff}_W$ can also
be used  in the Standard Model framework to obtain (a less precise)
$\Delta\alpha^{(5)}_{\rm 
had}(m^2_Z)$ which also disagrees with Eq.\ref{eq14}. Again, it seems
that one should lower the AY estimate of the high energy hadronic
contributions to \ahadp Taken together with the $\rho$ resonance region
reduction mentioned above, such a shift would then move AY's central
value down to 
almost exactly DH98b.  Of course, a full revision of AY should actually
be made in any serious update of their approach.

Based on the above comparisons, we conclude that although CLY85 and AY95
represent pioneering efforts to incorporate perturbative QCD into
$a_\mu$, their results are not state-of-the-art, are not well supported
by data and should not be seriously
considered on a par with more recent evaluations of \ahadp Furthermore,
an update of those analyses is likely to significantly reduce their
central values to the level very similar to more recent evaluations.

We should also note that several more recent (unpublished) updates or new studies of
\ahad beyond those in Table~1 have appeared. Eidelman and Jegerlehner
updated their 1995 result with newer data and found \cite{EJ99}
\beq
6967(119)\times10^{-11} \qquad ({\rm EJ99}) \label{eq15}
\eeqn
and Jegerlehner has further updated that result (from $e^+e^-$ data) to
\cite{J00} 
\beq
6974(105) \times 10^{-11} \qquad ({\rm J2000}) \label{eq16}
\eeqn

\noi A more recent theory-driven analysis (but with
parametrization set by $e^+e^-$ and $\tau$ decay data) due to Narison
found \cite{Na01}

\beq 
6970 (76) \times10^{-11} \qquad ({\rm N2001})  \label{eq17}
\eeqn

\noi All of these results are in good accord with DH98b
which was used in
comparison with $a^{\rm exp}_\mu$ \cite{1999}.
Any of the up-to-date studies will
give a deviation between 2 and 2.6 sigma when compared with experiment.

\section{Summary and Conclusions}

The QED and weak contributions to \g2 are known to an accuracy well below what
might be accessible to E821. The higher-order hadronic contribution 
also appears to be under control.  The recent use of perturbative
QCD along with $\tau$-decay 
data have greatly improved the uncertainty in the leading hadronic
contribution, and supersedes the earlier analyses which only 
used (more limited) 
data from electron-positron annihilation to hadrons as input.  
It is clear from the literature that one needs to be careful when using 
parameterizations in the important $\rho$ region, unless well supported
by data.  The use of perturbative QCD at relatively low energies,
which was pioneered by CLY85 has led 
to substantial gains in our understanding, and the agreement of the
recent BES data \cite{zz} with the QCD predictions of DH98a \cite{DH98a}
gives one confidence in the validity of the role of modern QCD studies in
this discussion.

The knowledge of the hadronic contribution to \am
has improved dramatically from the mid 1980s when the \g2 experiment began.
New high quality $e^+e^-$ data have become available from
Novosibirsk ($\sqrt{s} = $ threshold - 1.4 GeV), and from Bejing
($\sqrt{s} = $ 2.0 - 5.0 GeV).  The entire sample of LEP $\tau$-decay
data, as well as the CLEO  $\tau$ data \cite{cleo} have also become
available.  An updated global analysis of all of these data, along with QCD
information is underway by Eidelman, Davier and H\"ocker.  These
authors have now spent many years working on this topic, and we welcome
their combined efforts to produce a new value for \ahadp  This new more
complete 
analysis should make all of the previous analyses obsolete.

Currently the best published evaluation of \ahad is the work by Davier and 
H\"ocker \cite{DH98a,DH98b}.  We see no reason to ignore the substantial
amount of additional information which went into these analyses, and
rather to
use an earlier evaluation which contains only part of our current
knowledge.  To us, the assertion that one can  
ignore the latest information, or the claim that earlier,
and by necessity, less complete analyses are on an equal footing  is
incomprehensible, particularly when the earlier approach \cite{AY}
fails so badly 
for $\Delta \alpha_{\rm had} (m^2_Z)$ and has other serious
deficiencies. Both in experimental and 
theoretical physics,  
it has been the course of scientific progress
to have new information and improved analyses of problems 
replace the results of earlier efforts. 

We are fortunate that at this point several independent authors have
considerable experience 
in the delicate issues which are involved in obtaining a value for
\ahadp  They have brought  new insights to the field.  We anxiously
await their updated evaluation of this important number.

\bigskip
We wish to thank Michel Davier for helpful comments on a number of points.
LR wishes to thank J. Bijnens, D.H. Brown, T, Kinoshita
and E. de Raphael for useful discussions.  This work was supported in part
by the U.S. NSF and the U.S. DOE.

\newpage

\addtocounter{table}{-3}

\end{document}